# Disease Detectives: Using Mathematics to Forecast the Spread of Infectious Diseases

by Heather Z. Brooks, Unchitta Kanjanasaratool, Yacoub H. Kureh, and Mason A. Porter

**Abstract**: The COVID-19 pandemic has led to significant changes in how people are currently living their lives. To determine how to best reduce the effects of the pandemic and start reopening societies, governments have drawn insights from mathematical models of the spread of infectious diseases. In this article, we give an introduction to a family of mathematical models (called "compartmental models") and discuss how the results of analyzing these models influence government policies and human behavior, such as encouraging mask wearing and physical distancing to help slow the spread of the disease.

## Section 1: Modeling Infectious Diseases

At the end of 2019, doctors and scientists learned about a new virus, now called severe acute respiratory syndrome coronavirus 2 (SARS-CoV-2), that was spreading in China. The virus causes coronavirus disease 2019 (COVID-19) [1,4,11] and has spread throughout the world as a global pandemic. (See [12] to develop some intuition about pandemics and [1,8] for an accessible introduction to the COVID-19 pandemic.) What makes this virus so dangerous is how it spreads from person to person and the effects that it has on people who catch it.

Scientists play many roles in helping people recover from a virus through the design of medicine and medical equipment. They also research other ways — including with mathematics and computation — to keep people safe by studying the effects of actions such as physical distancing and wearing masks. Governments can then use such information to develop health guidelines and policies. In this article, we introduce the mathematical modeling of infectious diseases [3,9,10]. Scientists who do this are sometimes called *mathematical epidemiologists*.

To better understand how a disease spreads, scientists use a combination of mathematics and data (so-called "mathematical modeling"). The most common procedure is to use what is called a *compartmental model* [3,7], an approach for studying disease spread that was invented almost 100 years ago and has been used successfully for *forecasting* in many epidemics. Compartmental models provide a way to formulate simple rules that approximate how a virus like SARS-CoV-2 spreads. The models can behave in complicated ways even when the rules are simple. When creating and studying such models, one seeks to improve the accuracy of forecasting disease spread and to provide a way to test the effects of possible responses (for example, if everyone stays at home) for reducing the spread. These results can inform people about helpful guidelines, policy, and actions [3].

In a compartmental model of an epidemic (see Figure 1), one separates a population into categories (called "compartments") and examines how people change categories. A frequently studied type of compartmental model is an *SIR model*, where SIR is short for susceptible–infected–recovered. The "S" compartment consists of people who are susceptible to infection (this means that they can get the disease), and the "I" compartment consists of people who are infected (and who can infect others). The "R" compartment is for people who have recovered from the infection, although "R" can also stand for "removed" to account for people who die from an infection.

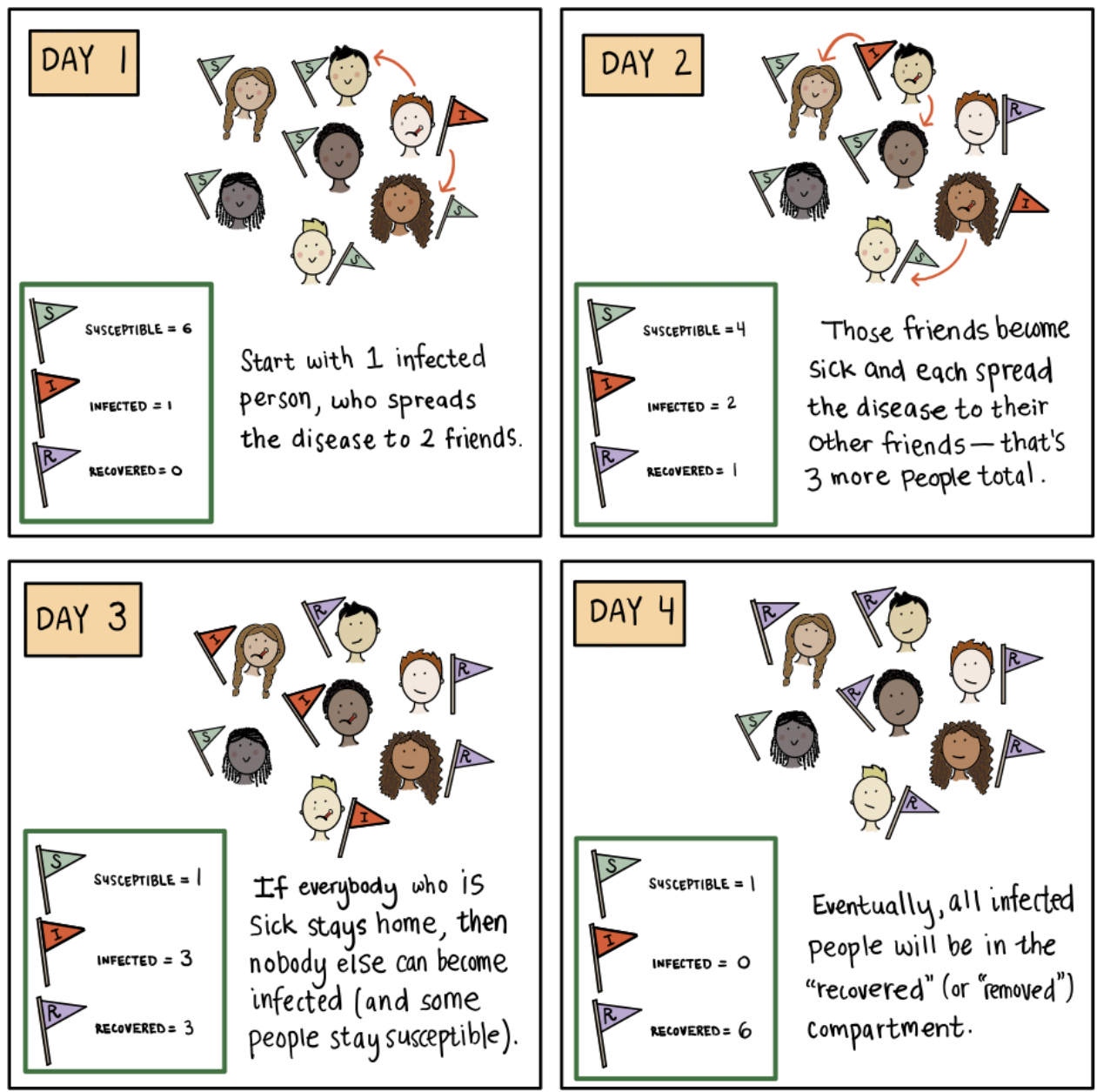

*Figure 1.* An illustration of a compartmental model of an infectious disease with susceptible (S), infected (I), and recovered (R) people. This type of model is known as an SIR model.

In developing an epidemic model, it is important to focus on the most important factors of a disease and the particular scientific questions of interest so that the model is as simple and informative as possible. These factors are different for different diseases. For example, one consideration in an epidemic model is social interactions (such as shaking hands, seeing a movie together, or playing board games together in someone's house) between people [7,9], as such social *contact* gives an opportunity for a virus to spread. Some types of disease require very close contact to spread, but others can spread even through indirect contact (such as from touching the same surface or just being nearby). For the COVID-19 pandemic, this is why people have been encouraged to practice "physical distancing" (often called "social distancing") and to wear masks.

Section 2: How Quickly Does a Disease Spread?

Let's illustrate the modeling of the spread of an infectious disease with an SIR model [3,7]. Before we can use such a model to forecast the effects of a disease on a population, we have to know (or estimate) a few important factors:
1. The amount of time that a person is infectious. This tells us how long they can infect other people.
2. The rate of contact in a population.
3. The chance that a contact leads to an infection.

In combination, these three factors allow one to estimate a quantity called the *basic reproduction number*. The basic reproduction number, denoted by $R_0$ (pronounced "R naught"), is the number of people, on average, to whom a single infected person spreads a disease if they are placed in a population of susceptible people. A quantity that is closely related to $R_0$ is the effective reproduction number, which is equal to $R_0$ multiplied by the fraction of a population who are susceptible. The effective reproduction number is important because some individuals in a population may be immune to a disease, as perhaps they were vaccinated or developed immunity from a past infection.

Suppose that we start with some population (such as people in a city, like Los Angeles). Before a disease begins to spread, everyone in the population is in the "susceptible" compartment. Now suppose that someone who has the disease flies into Los Angeles and starts to spread it to other people in the city (see Figure 1). If $R_0$ is 2 and the time that a person is infectious is 1 day, then, on average, that person will spread the disease to 2 other people before recovering, and those 2 people in turn will spread it to 2 more people each before recovering, and so on. In this simplified setting, one can estimate how many people will become infected based on transmission and recovery rates.

When $R_0$ is larger than 1 (which we write mathematically as "$R_0 > 1$"), the number of infected people grows *exponentially*. To see how this works, let's use our above example with $R_0 = 2$ and infectiousness time of 1 day. Let's also suppose that we're looking at how many new infections occur on each day. Consider the following scenario. Suppose that 1 person is infected on the first day and spreads the disease to 2 other people on the second day. (Remember that the first person is no longer infectious on the second day.) On the third day, those 2 infected people can each infect 2 more susceptible people. In 3 days, that implies that we expect to have about 1 x 2 x 2 = 4 infected people. If this pattern continues and there are still many susceptible people, can you see how many new infections we get on the 4th day? We multiply by 2 again, so we expect to have about 8 newly infected people on the next day. After we do this calculation, we have to remind ourselves that these infections come from just 1 infected person. If we instead start with 100 infected people, we can see how the situation can become very bad very quickly.

The number of infected people from a disease continues to increase until the rate at which infected people recover exceeds the rate at which they infect susceptible people. If each infected person infects fewer than 1 other person on average, we expect fewer infections each day and that the disease will die out. How long this takes, and whether it occurs in the first place, depends on the size of a population and the social contacts of the people in it. Try the interactive SIR model in [5]. You can also take a look at the discussions in [9,12] and the discussions and interactive simulations in [8].

In Figure 2, we compare what happens to susceptible, infectious, and recovered people in a population when $R_0 > 1$ versus $R_0 < 1$. The vertical axes in the plots indicate the number of people who are infected, and the horizontal axes indicate the number of days since the first person was infected. The values on the vertical axis include all of the people in the susceptible, infected, or recovered classes on that day (not just the infections or recoveries that are new that day). In the top row (for which $R_0 > 1$), almost everybody eventually becomes infected in this simplistic scenario, so the number of susceptible people (the green curve) becomes very small. How quickly a disease takes hold in a population depends on the value of $R_0$. When $R_0 > 1$, the number of people who are infected at the same time can be very large (the orange curve), and hospitals may not have the capacity to treat everybody who becomes infected. In the bottom row (for which $R_0 < 1$), the disease does not spread to many people in the population, so the curve of infected people (again in orange) over time is much flatter. This scenario is desirable, and if a disease is spreading quickly, we want to slow infections and "flatten the curve" (see Figure 3). We discuss "flattening the curve", along with behavior and policies that can help do this, in Section 3. Additionally, now that we have discussed $R_0$ some more, take another look at what happens when you change quantities like infection rates and recovery rates in the interactive simulation in [5].

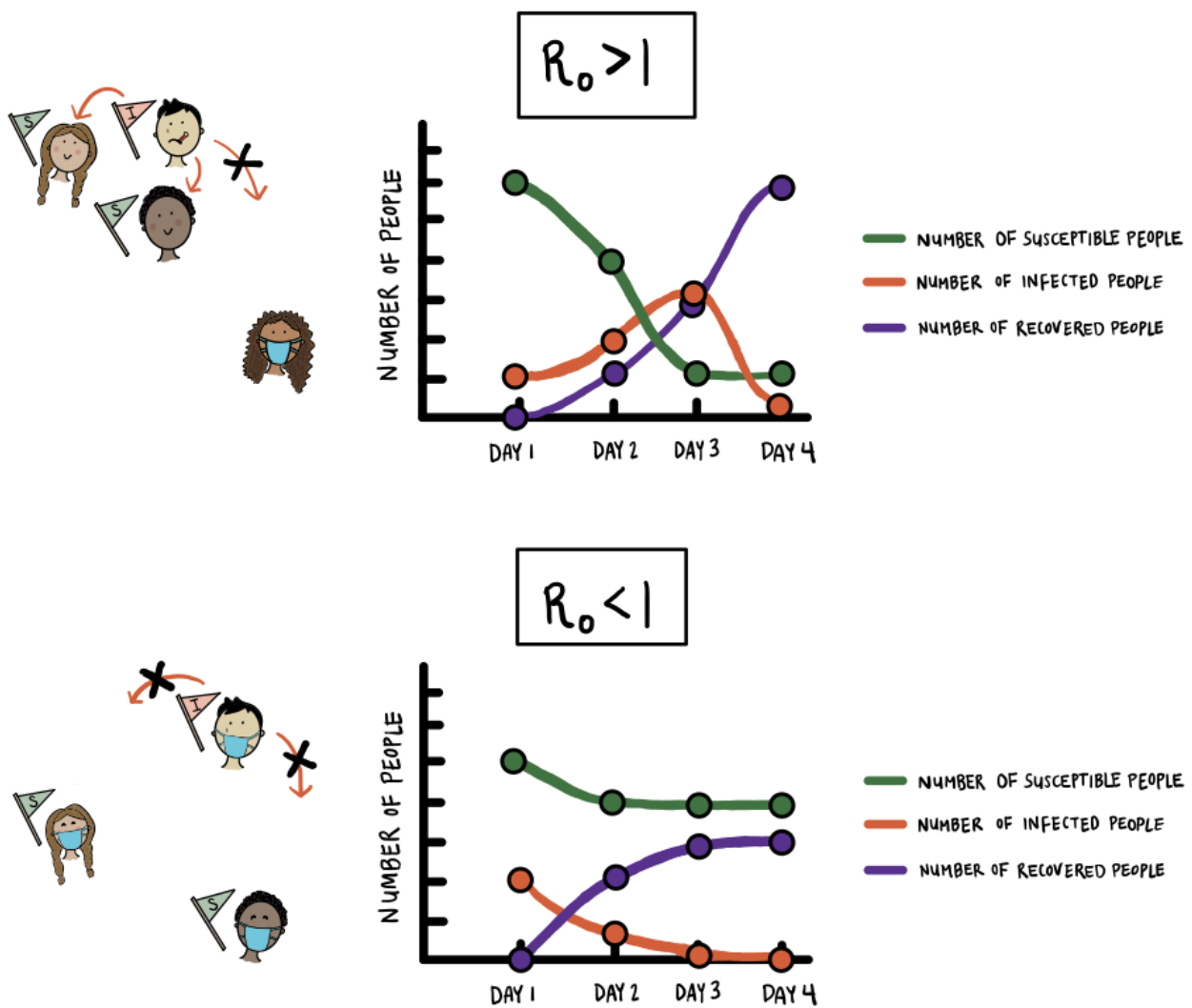

*Figure 2.* Comparison of the number of susceptible, infected, and recovered people in an SIR model of an infectious disease when the basic reproduction number $R_0$ is (top) larger than 1 and (bottom) smaller than 1. The bottom panel illustrates a situation in which many people are wearing masks and practicing physical distance.

# Section 3: "Flattening the curve", policy, and behavior

In Section 2, we explored how the basic reproduction number $R_0$ influences how a disease spreads. Networks of social contacts are important for estimating the value of $R_0$ [2,7,8,9]. You and the friends whom you see in person are connected to each other in a contact network, and so are your parents and their friends. Therefore, if one of your friends gets a disease, it is pretty likely to spread to you and then to your parents and then to their friends.

Because of how easily people can become infected (and perhaps very sick) from COVID-19, many governments responded by closing schools, canceling sporting events and other large gatherings of people, quarantining infected people, and telling people to stay at home and practice physical distancing. The goals of these kinds of policies and behaviors during a pandemic [8,9,12] is to try to limit both direct and indirect contacts and thereby "flatten the curve" of the number of infected people (see Figures 2 and 3). With measures like these, along with wearing masks and washing one's hands often, the orange curve of infected people in Figure 2 spreads out more over time and has a lower peak, indicating that the maximum number of infected people on any one day is smaller than it would otherwise be. This is important for hospitals to have enough capacity to treat as many people as possible, as indicated by the heights of the peak above the dashed horizontal lines in Figure 3. With a flattened curve, although a disease continues to spread, there is more space in hospitals to treat infected people who need help; this reduces the number of deaths from a disease. Flattening the curve also decreases the total number of people who become infected over time (these are the colored regions of Figure 3).

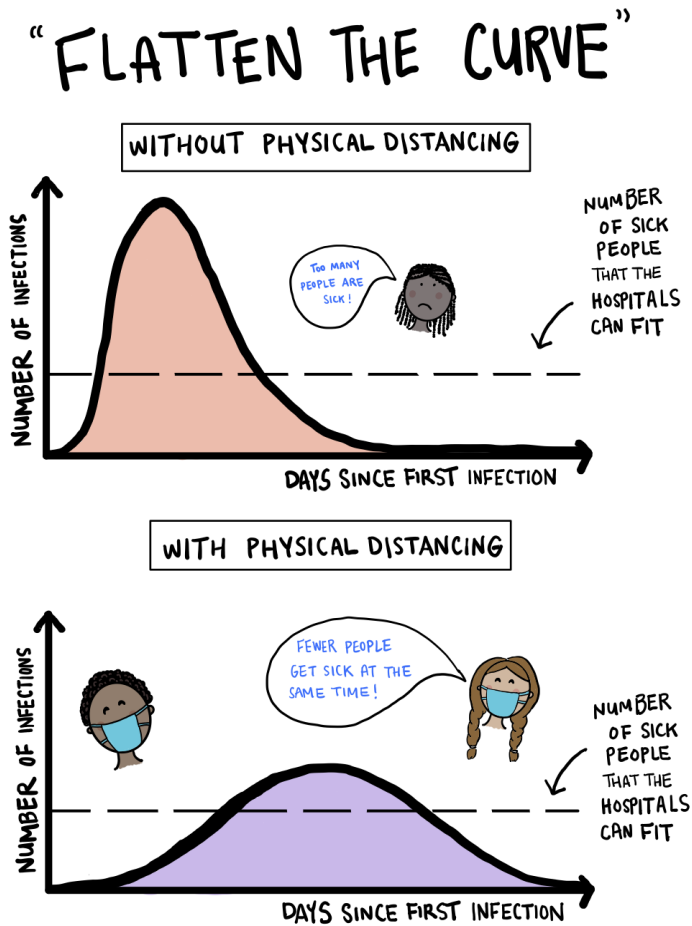

*Figure 3. Illustration of "flattening the curve" of infections by using physical distancing (often called "social distancing"). [Our figure is inspired by pictures such as the one at https://commons.wikimedia.org/wiki/File:Covid-19-curves-graphic-social-v3.gif.]*

## Section 4: Conclusions

Mathematical models and computation have played a major role in influencing the responses of governments to the COVID-19 pandemic. These models — such as the one in [2], which influenced policy in Spain, and the one in [6], which influenced policy in the United Kingdom — are much more detailed than the SIR model that we discussed above. For example, some models include compartments for hospitalization and for people without symptoms who can infect others. Other important data to include are contact-network structure (as we discussed above) and ages and other demographic factors (which may influence recovery and death rates).

The current COVID-19 pandemic illustrates the importance of the mathematical modeling of infectious diseases. Mathematical and computational approaches allow people to make progress when laboratory experiments are impractical or impossible, and they also allow one to get at the core of a difficult problem and develop ideas for how to solve it.

## Glossary

**Basic reproduction number ($R_0$).** The number of cases of infection, on average, that are generated by a single infected person in a population of susceptible people.

**Compartment.** A category in a compartmental model of epidemics. For example, in a susceptible–infected–recovered (SIR) model, there are compartments for susceptible people (who can become infected), infected people, and recovered people.

**Compartmental model of an epidemic.** An epidemic model is a set of rules (typically in mathematical form) that describe how the properties of an epidemic change in time. A common type of epidemic model is a compartmental model, in which each type of compartment is a category of individuals. In such a model, there are rules that determine how people change categories. An example of a compartmental model is a susceptible–infected–recovered (SIR) model, which includes equations to describe how the numbers of susceptible, infected, and recovered people change over time.

**Contact.** A connection between two individuals. There are both direct contacts (such as shaking hands) and indirect contacts (such as touching the same surface or just being nearby). In a contact network, each connection between individuals in a network represents some type of contact.

**Epidemiologist.** An epidemiologist is a scientist who studies diseases, including the spread of diseases and how to control them.

**Exponential growth.** A particularly fast type of growth in which the rate of growth of something at any particular time is proportional to the current amount of that thing. For example, if the number of infected people triples every 2 days, then if there is 1 infected person on the first day, there will be

3 infected people on the third day, 9 infected people on the fifth day, 27 infected people on seventh day, 81 infected people on ninth day, and so on. If the disease keeps spreading this fast, there would be 19683 infected people on the nineteenth day.

**Forecast.** A form of prediction in which one indicates a range of possibilities (rather than an exact number) of future events, such as a 42% chance of rain in Los Angeles tomorrow or a range of the number of people in Los Angeles who will become infected with COVID-19 during December 2020.

**Mathematical epidemiology.** The study of the mathematics of infectious diseases, such as through the mathematical modeling of such diseases.

**Pandemic.** An epidemic that has spread across a large region, such as worldwide or across several continents. See [12] to develop some intuition about pandemics and [9] for a discussion of modeling disease spread in a pandemic. See [8] for an accessible introduction to the COVID-19 pandemic.

## Conflict of Interest statement
The authors declare that the research was conducted in the absence of any commercial or financial relationships that could be construed as a potential conflict of interest.

## Acknowledgements
We are grateful to our young readers — Emily Chen, Nia Chiou, Taryn Chiou, Dimitri Chrysafis, Maria Chrysafis, Valerie K. Eng, Iris Leung, Talan Li, Adam Lindemood, Suzanna Lindemood, Eli Truong — for their many helpful comments. We also thank their parents, teachers, and friends — Alena Carter, Lyndie Chiou, and Christina Chow — for putting us in touch with them and soliciting their feedback. Additionally, we thank John Butler, Francesca Henderson, Rachel Levy, Joel Miller, and our referees for helpful comments. MAP acknowledges support from the National Science Foundation (grant number DMS-2027438) through the RAPID program; and MAP and YHK acknowledge support from the National Science Foundation (grant number 1922952) through the Algorithms for Threat Detection (ATD) program.

## Author Biographies
**Heather Zinn Brooks** was born in Idaho and grew up in Salt Lake City, Utah. Heather got her Ph.D. in Mathematics from the University of Utah in 2018. Now, she lives in sunny Los Angeles, where she is an assistant professor at Harvey Mudd College. In her work, Heather focuses on mathematical modeling of real-world applications, including the spreading of diseases and misinformation. Outside of work, Heather is spending her quarantine doing puzzles, listening to music, and making art. Once physical distancing is over, she's especially excited to go climbing and to share good food with family and friends.

**Unchitta Kanjanasaratool** was born and raised in Bangkok, Thailand before she moved to California in 2016. Unchitta finished her undergraduate degree in Applied Mathematics at UCLA in 2020 and is now pursuing a Ph.D. in Computational Social Science at George Mason University. Unchitta gets excited by cool applications of mathematics, especially in social sciences, and enjoys communicating it with others. She looks forward to exploring more of the Washington D.C. area where she recently moved to, as well as playing ultimate frisbee (her favorite sport) with her friends again once the COVID-19 pandemic is under control.

**Yacoub H. Kureh** was born and raised in Orange County, California and earned his Ph.D. in Mathematics from UCLA in 2020. His research interests are in the areas of network science and data science. He focuses on modeling opinions spreading among individuals who share social connections. He is a first-generation college and graduate student. He is passionate about education and supporting the right to education. While in quarantine, he spends his time reading science fiction, cycling, and playing online board games with his friends. He is looking forward to going backpacking with friends when physical-distancing orders are lifted.

**Mason A. Porter** is a professor in the Department of Mathematics at UCLA. He was born in Los Angeles, California, and he is excited to be a professor in his hometown. In addition to studying networks and other topics in mathematics and its applications, Mason is a big fan of games of all kinds, fantasy, baseball, the 1980s, and other delightful things. Mason used to be a professor at University of Oxford, where he did actually wear robes on occasion (like in the *Harry Potter* series). He is very eager to stop physical distancing (although he is determined to do it as long as necessary) and see his friends and students again in person, and he looks forward to spending quality time with them.